\documentstyle[12pt,psfig,leqno]{article}
\topmargin - 1.0 true cm
\def\ss{\footnotesize}
\def\SS{\footnotesize}
\def\sss{\scriptscriptstyle}
\def\barp{{\raise.35ex\hbox{${\sss (}$}}---{\raise.35ex\hbox{${\sss )}$}}}
\def\bdbarp{\hbox{$B_d$\kern-1.4em\raise1.4ex\hbox{\barp}}}
\def\bsbarp{\hbox{$B_s$\kern-1.4em\raise1.4ex\hbox{\barp}}}
\def\dbarp{\hbox{$D$\kern-1.1em\raise1.4ex\hbox{\barp}}}
\def\dcp{D^0_{\sss CP}}
\def\dbar{{\overline{D^0}}}
\newcommand{\xd}{x_d}
\newcommand{\xs}{x_s}
\newcommand{\bd}{B_d^0}
\newcommand{\bdb}{\overline{B_d^0}}
\newcommand{\bs}{B_s^0}
\newcommand{\bsb}{\overline{B_s^0}}
\newcommand{\bu}{B_u^\pm}
\newcommand{\beq}{\begin{equation}}
\newcommand{\eeq}{\end{equation}}
\newcommand{\absvcb}{\vert V_{cb}\vert}
\newcommand{\absvub}{\vert V_{ub}\vert}

\newcommand{\abseps}{\vert\epsilon\vert}

\newcommand{\fbb}{f^2_{B_d}\hat{B}_{B_d}}
\newcommand{\fbbs}{f^2_{B_s}\hat{B}_{B_s}}
\newcommand{\fbd}{f_{B_d}}
\newcommand{\fbs}{f_{B_s}}

\def\rly#1{\mathrel{\raise.3ex\hbox{$#1$\kern-.75em\lower1ex\hbox{$\sim$}}}}
\def\lsim{\rly<}

\def \zpc#1#2#3{{\it Z.~Phys.,} C#1 (19#2) #3}
\def \plb#1#2#3{{\it Phys.~Lett.,} B#1 (19#2) #3}
\def \ibj#1#2#3{~#1, (19#2) #3}
\def \prl#1#2#3{{\it Phys.~Rev.~Lett.,} #1 (19#2) #3}
\def \prd#1#2#3{{\it Phys.~Rev.,} D#1 (19#2) #3}
\def \npb#1#2#3{{\it Nucl.~Phys.}, B#1 (19#2) #3}

\newread\epsffilein 
\newif\ifepsffileok 
\newif\ifepsfbbfound 
\newif\ifepsfverbose 
\newdimen\epsfxsize 
\newdimen\epsfysize 
\newdimen\epsftsize 
\newdimen\epsfrsize 
\newdimen\epsftmp  
\newdimen\pspoints  
\def\epsfbox#1{\global\def\epsfllx{72}\global\def\epsflly{72}%
 \global\def\epsfurx{540}\global\def\epsfury{720}%
 \def\lbracket{[}\def\testit{#1}\ifx\testit\lbracket
 \let\next=\epsfgetlitbb\else\let\next=\epsfnormal\fi\next{#1}}%
\def\epsfgetlitbb#1#2 #3 #4 #5]#6{\epsfgrab #2 #3 #4 #5 .\\%
 \epsfsetgraph{#6}}%
\def\epsfnormal#1{\epsfgetbb{#1}\epsfsetgraph{#1}}%
\def\epsfgetbb#1{%
\openin\epsffilein=#1
\ifeof\epsffilein\errmessage{I couldn't open #1, will ignore it}\else
 {\epsffileoktrue \chardef\other=12
 \def\do##1{\catcode`##1=\other}\dospecials \catcode`\ =10
 \loop
  \read\epsffilein to \epsffileline
  \ifeof\epsffilein\epsffileokfalse\else
   \expandafter\epsfaux\epsffileline:. \\%
  \fi
 \ifepsffileok\repeat
 \ifepsfbbfound\else
 \ifepsfverbose\message{No bounding box comment in #1; using defaults}\fi\fi
 }\closein\epsffilein\fi}%
\def\epsfclipstring{}
\def\epsfsetgraph#1{%
 \epsfrsize=\epsfury\pspoints
 \advance\epsfrsize by-\epsflly\pspoints
 \epsftsize=\epsfurx\pspoints
 \advance\epsftsize by-\epsfllx\pspoints
 \epsfxsize\epsfsize\epsftsize\epsfrsize
 \ifnum\epsfxsize=0 \ifnum\epsfysize=0
  \epsfxsize=\epsftsize \epsfysize=\epsfrsize
  \epsfrsize=0pt
  \else\epsftmp=\epsftsize \divide\epsftmp\epsfrsize
  \epsfxsize=\epsfysize \multiply\epsfxsize\epsftmp
  \multiply\epsftmp\epsfrsize \advance\epsftsize-\epsftmp
  \epsftmp=\epsfysize
  \loop \advance\epsftsize\epsftsize \divide\epsftmp 2
  \ifnum\epsftmp>0
   \ifnum\epsftsize<\epsfrsize\else
    \advance\epsftsize-\epsfrsize \advance\epsfxsize\epsftmp \fi
  \repeat
  \epsfrsize=0pt
  \fi
 \else \ifnum\epsfysize=0
  \epsftmp=\epsfrsize \divide\epsftmp\epsftsize
  \epsfysize=\epsfxsize \multiply\epsfysize\epsftmp
  \multiply\epsftmp\epsftsize \advance\epsfrsize-\epsftmp
  \epsftmp=\epsfxsize
  \loop \advance\epsfrsize\epsfrsize \divide\epsftmp 2
  \ifnum\epsftmp>0
  \ifnum\epsfrsize<\epsftsize\else
   \advance\epsfrsize-\epsftsize \advance\epsfysize\epsftmp \fi
  \repeat
  \epsfrsize=0pt
 \else
  \epsfrsize=\epsfysize
 \fi
 \fi
 \ifepsfverbose\message{#1: width=\the\epsfxsize, height=\the\epsfysize}\fi
 \epsftmp=10\epsfxsize \divide\epsftmp\pspoints
 \vbox to\epsfysize{\vfil\hbox to\epsfxsize{%
  \ifnum\epsfrsize=0\relax
  \includegraphics{#1}%
  \else
  \epsfrsize=10\epsfysize \divide\epsfrsize\pspoints
  \includegraphics{#1}%
  \fi
  \hfil}}%
\global\epsfxsize=0pt\global\epsfysize=0pt}%
 {\catcode`\%=12 \global\let\epsfpercent=
\long\def\epsfaux#1#2:#3\\{\ifx#1\epsfpercent
 \def\testit{#2}\ifx\testit\epsfbblit
  \epsfgrab #3 . . . \\%
  \epsffileokfalse
  \global\epsfbbfoundtrue
 \fi\else\ifx#1\par\else\epsffileokfalse\fi\fi}%
\def\epsfempty{}%
\def\epsfgrab #1 #2 #3 #4 #5\\{%
\global\def\epsfllx{#1}\ifx\epsfllx\epsfempty
  \epsfgrab #2 #3 #4 #5 .\\\else
 \global\def\epsflly{#2}%
 \global\def\epsfurx{#3}\global\def\epsfury{#4}\fi}%
\def\epsfsize#1#2{\epsfxsize}
\def\mt{m_t}
\def\mb{m_b}
\def\mc{m_c}

\newcommand{\delmd}{\Delta M_d}
\newcommand{\delms}{\Delta M_s}

\newcommand{\kkbar}{$K^0$-${\overline{K^0}}$}
\newcommand{\bdbdbar}{$B_d^0$-${\overline{B_d^0}}$}
\newcommand{\bsbsbar}{$B_s^0$-${\overline{B_s^0}}$}

\begin{document}
\begin{flushright}
DESY 95-148 \\
UdeM-GPP-TH-95-32\\
\end{flushright}
\begin{center}
{\Large \bf
\centerline
{CP Violation and Flavour Mixing in the Standard Model}}
\vspace*{1.5cm}
 {\large A.~Ali}$\footnote{Presented at the 6th. International
Symposium on Heavy Flavour Physics, Pisa, June 6 - 10, 1995.}$
\vskip0.2cm
  Deutsches Elektronen Synchrotron DESY, Hamburg \\
\vspace*{0.3cm}
\centerline{ and}
\vspace*{0.3cm}
\centerline{\large D.~London}
\smallskip
  Laboratoire de physique nucl\'eaire, Universit\'e de
Montr\'eal \\
    C.P. 6128, succ. centre-ville, Montr\'eal, QC, Canada
H3C 3J7\\
\vskip0.5cm
{\Large Abstract\\}
\parbox[t]{\textwidth}{
\indent
We review and update the constraints on the parameters of the quark flavour
mixing matrix $V_{CKM}$ in the standard model and estimate the resulting CP
asymmetries in $B$ decays, taking into account recent experimental and
theoretical developments. In performing our fits, we use inputs from the
measurements of the following quantities: (i) $\abseps$, the CP-violating
parameter in $K$ decays, (ii) $\delmd$, the mass difference due to the
\bdbdbar\ mixing, (iii) the matrix elements $\absvcb$ and $\absvub$, (iv)
$B$-hadron lifetimes, and (v) the top quark mass. The experimental input in
points (ii) - (v) has improved compared to our previous fits. With the
updated CKM matrix we present the currently-allowed range of the ratios
$|V_{td}/V_{ts}|$ and $|V_{td}/V_{ub}|$, as well as the standard model
predictions for the \bsbsbar\ mixing parameter $\xs$ (or, equivalently,
$\delms$) and the quantities $\sin 2\alpha$, $\sin 2\beta$ and
$\sin^2\gamma$, which characterize the CP-asymmetries in $B$-decays.
Various theoretical issues related to the so-called ``penguin-pollution,"
are of importance for the determination of the phases $\alpha$ and $\gamma$
from the CP-asymmetries in $B$ decays, are also discussed.
}
\end{center}
\thispagestyle{empty}
\newpage
\setcounter{page}{1}
\textheight 23.0 true cm


\section{Introduction}

The aim of this article is to revise and update the profile of the
Cabibbo-Kobayashi-Maskawa (CKM) matrix reported earlier by us \cite{AL94},
in particular the CKM unitarity triangle and the CP asymmetries in $B$
decays, which are the principal objects of interest in experiments at
present and  forthcoming $B$ facilities. In performing this update, we
include the improvements reported in a number of measurements of the
lifetime, mixing ratio, and the CKM matrix elements $\absvcb$ and $\vert
V_{ub}/V_{cb} \vert$ from $B$ decays, as well as the top quark mass. On the
theoretical side, we mention the improved estimates of the power
corrections in the analysis of the exclusive semileptonic decay $B \to D^*
\ell \nu_\ell$ in the context of the heavy quark effective theory (HQET)
\cite{neubert95,shifman95}, and the calculation of the missing part of the
next-to-leading order calculations in the analysis of the CP-violating
quantity $\abseps$ \cite{HN95}. We note here the changes that we have made
in the input to our present analysis compared to that reported by us in
Ref.~\cite{AL94}:
\begin{itemize}

\item
The top quark (pole) mass $\mt =174 \pm 16$ GeV, measured earlier by the
CDF collaboration \cite{CDFmtold}, is now replaced by improved measurements
by the same collaboration \cite{CDFmtnew} and by D0 \cite{D0mtnew},
yielding the present world average $\mt =180 \pm 11 $ GeV
\cite{Belletini95}. This leads to the running top quark mass in the
$\overline{MS}$ scheme, $\overline{\mt} (\mt)= 170 \pm 11 $ GeV
\cite{mtmsbar}.

\item
A new and improved measurement of the quantity ${\cal F}(1) \vert V_{cb}
\vert$ in the decays $B \to D^* \ell \nu_\ell$, using methods based on the
heavy quark effective theory (HQET), has been reported by the ALEPH
collaboration \cite{ALEPHVcb95}. This is lower than their previous number
\cite{ALEPHVcb94}, as well as the corresponding numbers from the CLEO
\cite{CLEOVcb94} and ARGUS \cite{ARGUSVcb94} analyses. Likewise, new
measurements are reported by the DELPHI collaboration \cite{DELPHIVcb95}.
In the meantime, estimates for the quantity ${\cal F}(1) \equiv \xi(1)
\eta_A$ have undergone some revision in both the QCD perturbative part
$\eta_A$ and power corrections to the Isgur-Wise function at the symmetry
point $\xi(1)$. We use the value ${\cal F}(1) = 0.91\pm 0.04$, obtained
recently by Neubert \cite{neubert95}, and which is in good agreement with
the estimates of Shifman et al. \cite{shifman95,suv94}. Taking into account
the updated experimental and theoretical input, we obtain $\vert V_{cb}
\vert =0.0388 \pm 0.0036$. The central value for this matrix element has
come down compared to the value $\vert V_{cb} \vert =0.041 \pm 0.006$ used
by us previously, and the error on this quantity is now smaller, about $\pm
9\%$.

\item
Until recently, the knowledge of the CKM matrix element ratio  $\vert
V_{ub}/V_{cb} \vert$ was based on the analysis of the end-point lepton
energy spectrum in semileptonic $B$ decays \cite{Patterson}, which is quite
model-dependent. We had used a value  $\vert V_{ub}/V_{cb} \vert = 0.08 \pm
0.03$ to take this model dependence into account. In the meantime,
the measurement of the exclusive semileptonic decays $B \to (\pi, \rho) \ell
\nu_\ell$ has been reported by the CLEO collaboration \cite{Thorn95}.
The matrix element ratio so determined is also model-dependent due to the
decay form factors. However, this set of data permits a discrimination
among a number of models, all of which were previously  allowed from the
inclusive decay analysis. The convolution of the two methods reduces the
theoretical uncertainty somewhat. We use a value $\vert V_{ub}/V_{cb} \vert
= 0.08 \pm 0.02$, which is a fair reflection of the underlying present
theoretical dispersion on this ratio.

\item
In the analysis of the CP-violating quantity $\abseps$, the perturbative
renormalizations of the various pieces in the $\vert \Delta S \vert =2$
Hamiltonian from the intermediate charm and top quark are required
\cite{Burastop}. While the next-to-leading order results for the quantities
$\hat{\eta}_{cc}$ and $\hat{\eta}_{tt}$ have been known for some time and
were used in our previous analysis, the next-to-leading order calculation
of the quantity $\hat{\eta}_{ct}$ has been completed only recently
\cite{HN95}. We use the improved calculation of $\hat{\eta}_{ct}$ in
performing the CKM fits presented here.

\item
The measurements of the  $\bd$-$\bdb$ mass difference $\Delta M_d$ have
become quite precise,  with the present world average being $\Delta M_d
=0.465 \pm 0.024~(ps)^{-1}$ \cite{Wells}. The present lower limit on the
$\bs$-$\bsb$ mass difference has improved slightly, $\delms >
6.1~(ps)^{-1}$ at 95 \% C.L., assuming for the probability of the
fragmentation of a $b$ quark into a $B_s$ meson a value $f_s=12\%$
 \cite{ALEPHxs}, yielding  $\Delta M_s/\Delta M_d
> 12.3$ at $95 \%$ C.L. \cite{Wells}.

\end{itemize}
All of these improvements warrant an updated fit of the CKM parameters.

As in our previous analysis,  we consider two types of fits. In Fit 1, we
assume particular fixed values for the theoretical hadronic quantities. The
allowed ranges for the CKM parameters are derived from the (Gaussian)
errors on experimental measurements only. In Fit 2, we assign a central
value plus an error (treated as Gaussian) to the theoretical quantities. In
the resulting fits, we combine the experimental and theoretical errors in
quadrature. For both fits we calculate the allowed region in CKM parameter
space at 95\% C.L. We also estimate the SM prediction for the \bsbsbar\
mixing parameter, $\xs$, and show how the ALEPH limit of $\delms /\delmd >
12.3$ (95\% C.L.) constrains the CKM parameter space. We give the present
$95\%$ C.L. upper and lower bounds on the matrix element ratio $\vert
V_{td}/V_{ts} \vert$, as well as the allowed (correlated) values of the CKM
matrix elements $|V_{td}|$ and $|V_{ub}|$.

We also present the corresponding allowed ranges for the CP-violating
phases that will be measured in $B$ decays, characterized by $\sin 2\beta$,
$\sin 2\alpha$ and $\sin^2\gamma$. These can be measured directly through
rate asymmetries in the decays $\bdbarp \to J/\psi K_S$, $\bdbarp \to \pi^+
\pi^-$, and $\bsbarp\ \to D_s^\pm K^\mp$ (or $B^\pm \to \dbarp\ K^\pm$),
respectively. We also give the allowed domains for two of the angles,
$(\sin 2\alpha,\sin 2\beta)$. Finally, we briefly discuss the role of
penguins (strong and electroweak) in extracting the CP phases $\alpha$ and
$\gamma$ from the measurements of various CP-asymmetries.

This paper is organized as follows. In Section 2, we present our update of
the CKM matrix, concentrating especially on the matrix element $\absvcb$
which, thanks to the progress in HQET and experiments, is now well under
control. The constraints that follow from $\vert V_{ub}/V_{cb} \vert$,
$\abseps$ and $\Delta M_d$ on the CKM parameters are also discussed here.
Section 3 contains the results of our fits. These results are summarized in
terms of the allowed domains of the unitarity triangle, which are displayed
in several figures and tables. In Section 4, we discuss the impact of the
recent lower limit on the ratio $\Delta M_s/\Delta M_d$ reported by the
ALEPH collaboration on the CKM parameters and estimate the expected range
of the mixing ratio $\xs$ in the SM based on our fits. Here we also present
the allowed 95\% C.L. range for $\vert V_{td}/V_{ts} \vert$. In Section 5
we discuss the predictions for the CP asymmetries in the neutral $B$ meson
sector and calculate the correlations for the CP violating asymmetries
proportional to $\sin 2\alpha$, $\sin 2 \beta$ and $\sin^2 \gamma$. We also
review some of the possible theoretical uncertainties in extracting these
CP-phases due to the presence of strong and/or electroweak penguins. We
present here the allowed values of the CKM matrix elements $|V_{td}|$ and
$|V_{ub}|$. Section 6 contains a summary and an outlook for improving the
profile of the CKM unitarity triangle.


\section{An Update of the CKM Matrix}

In updating the CKM matrix elements, we make use of the Wolfenstein
parametrization \cite{Wolfenstein}, which follows from the observation that
the elements of this matrix exhibit a hierarchy in terms of $\lambda$, the
Cabibbo angle. In this parametrization the CKM matrix can be written
approximately as
\beq
V_{CKM} \simeq \left(\matrix{
 1-{1\over 2}\lambda^2 & \lambda
   & A\lambda^3 \left( \rho - i\eta \right) \cr
 -\lambda ( 1 + i A^2 \lambda^4 \eta )
& 1-{1\over 2}\lambda^2 & A\lambda^2 \cr
 A\lambda^3\left(1 - \rho - i \eta\right) & -A\lambda^2 & 1 \cr}\right)~.
\label{CKM}
\eeq
In this section we shall discuss those quantities which constrain these CKM
parameters, pointing out the significant changes in the determination of
$\lambda$, $A$, $\rho$ and $\eta$.

We recall that $\vert V_{us}\vert$ has been extracted with good accuracy
from $K\to\pi e\nu$ and hyperon decays \cite{PDG} to be
\beq
\vert V_{us}\vert=\lambda=0.2205\pm 0.0018~.
\eeq
This agrees quite well with the determination of $V_{ud}\simeq 1-{1\over
2}\lambda^2$ from $\beta$-decay,
\beq
\vert V_{ud}\vert=0.9744\pm 0.0010~.
\eeq

The parameter $A$ is related to the CKM matrix element $V_{cb}$, which can
be obtained from semileptonic decays of $B$ mesons. We shall restrict
ourselves to the methods based on HQET to calculate the exclusive and
inclusive semileptonic decay rates. In the heavy quark limit it has been
observed that all hadronic form factors in the semileptonic decays $B \to
(D,D^*) \ell \nu_\ell$ can be expressed in terms of a single function, the
Isgur-Wise function \cite{Wisgur}. It has been shown that the HQET-based
method works best for $B\to D^*l\nu$ decays, since these are unaffected by
$1/m_Q$ corrections \cite{Luke,Boyd,Neubert}. This method has been used by
the ALEPH, ARGUS, CLEO and DELPHI collaborations to determine $\xi (1)
\absvcb$ and the slope of the Isgur-Wise function.

Using HQET, the differential decay rate in $B \to D^* \ell \nu_\ell$ is
\begin{eqnarray}
\frac{d\Gamma (B \to D^* \ell \bar{\nu})}{d\omega }
&=& \frac{G_F^2}{48 \pi^3} (m_B-m_{D^*})^2 m_{D^*}^3 \eta_{A}^2
 \sqrt{\omega^2-1} (\omega + 1)^2 \\ \nonumber
&~& ~~~~~~~~~~~~~~\times [ 1+ \frac{4 \omega}{\omega + 1}
 \frac{1-2\omega r + r^2}{(1-r)^2}] \absvcb ^2 \xi^2(\omega) ~,
\label{bdstara1}
\end{eqnarray}
where $r=m_{D^*}/m_B$, $\omega=v\cdot v'$ ($v$ and $v'$ are the
four-velocities of the $B$ and $D^*$ meson, respectively), and $\eta_{A}$
is the short-distance correction to the axial vector form factor. In the
leading logarithmic approximation, this was calculated by Shifman and
Voloshin some time ago -- the so-called hybrid anomalous dimension
\cite{hybrid}. In the absence of any power corrections, $\xi (\omega=1)=1$.
The size of the $O(1/\mb^2)$ and $O(1/\mc^2)$ corrections to the Isgur-Wise
function, $\xi (\omega)$, and partial next-to-leading order
corrections to $\eta_A$ have received a great deal of theoretical
attention, and the state of the art has been
summarized recently by Neubert \cite{neubert95} and Shifman
 \cite{shifman95}. Following them, we take:
\begin{eqnarray}
\label{neubertxiold}
\xi (1) &=& 1+ \delta (1/m^2)= 0.945 \pm 0.025 ~, \nonumber \\
 \eta_{A} &=& 0.965 \pm 0.020 ~.
\end{eqnarray}
This gives the range \cite{neubert95}:
\beq
{\cal F}(1)=0.91 \pm 0.04~.
\label{alxi}
\eeq
The present experimental input from the exclusive
semileptonic channels is based on the data by
 CLEO \cite{CLEOVcb94}, ALEPH
\cite{ALEPHVcb95}, ARGUS \cite{ARGUSVcb94}, and DELPHI \cite{DELPHIVcb95}:
\begin{eqnarray}
\label{Vcbf195}
 \vert V_{cb}\vert \cdot {\cal F}(1)
  &=& 0.0351 \pm 0.0019 \pm 0.0020 ~~~[\mbox{CLEO}], \nonumber \\
  &=& 0.0314 \pm 0.0023 \pm 0.0025 ~~~[\mbox{ALEPH}], \nonumber \\
  &=& 0.0388 \pm 0.0043 \pm 0.0025 ~~~[\mbox{ARGUS}], \nonumber \\
  &=& 0.0374 \pm 0.0021 \pm 0.0034 ~~~[\mbox{DELPHI}],
\end{eqnarray}
 where the first error is statistical and the second systematic.
 The ARGUS number has been updated by taking
into
account the updated lifetimes for the $B^0$ and $B^\pm$ mesons
\cite{komamiya95}.
 The statistically weighted average of these numbers is:
\beq
  \vert V_{cb}\vert \cdot {\cal F}(1) = 0.0353 \pm 0.0018 ~,
\eeq
 which, using ${\cal F}(1)$ from Eq.~(\ref{alxi}), gives the following
value:
 \beq
  \vert V_{cb} \vert= 0.0388 \pm 0.0019 ~(expt) \pm 0.0017 ~(th).
\label{Vcbhqet95}
\eeq
Combining the errors linearly gives $\vert V_{cb} \vert = 0.0388 \pm
0.0036$. This is in good agreement with the value $\absvcb = 0.037
^{+0.003}_{-0.002}$ obtained from the exclusive decay $B \to D^* \ell
\nu_\ell$, using a dispersion relation approach \cite{BGL95}.
 Likewise, the value of $\vert V_{cb}
\vert$ obtained from the inclusive semileptonic $B$
 decays using HQET is quite compatible
with the above determination \cite{Patterson}:
\beq
 \vert V_{cb} \vert= 0.040 \pm 0.0010 ~(expt) \pm 0.005 ~(th) ~.
\eeq
 In the fits below we shall use
 $ \vert V_{cb} \vert = 0.0388 \pm 0.0036$, yielding
\beq
A = 0.80 \pm 0.075~.
\label{Avalue}
\eeq

The other two CKM parameters $\rho$ and $\eta$ are constrained by the
measurements of $\vert V_{ub}/V_{cb}\vert$, $\abseps$ (the CP-violating
parameter in the kaon system), $\xd$ (\bdbdbar\ mixing) and (in principle)
$\epsilon^\prime/\epsilon$ ($\Delta S=1$ CP-violation in the kaon system).
We shall not discuss the constraints from $\epsilon^\prime/\epsilon$, due
to the various experimental and theoretical uncertainties surrounding it at
present, but take up the rest in turn and present fits in which the allowed
region of $\rho$ and $\eta$ is shown.

Up to now, $\vert V_{ub}/V_{cb}\vert$ was obtained by looking at the
endpoint of the inclusive lepton spectrum in semileptonic $B$ decays.
Unfortunately, there still exists quite a bit of model dependence in the
interpretation of the inclusive data by themselves. As mentioned earlier, a
recent new input to this quantity is provided by the measurements of the
exclusive semileptonic decays $B \to (\pi, \rho) \ell \nu_\ell$. The ratios
of the exclusive semileptonic branching ratios provide some discrimination
among the various models \cite{Thorn95}. In particular, models such as that
of Isgur et al.~\cite{ISGW}, which give values in excess of 3 for the
ratio of the decay widths $\Gamma (B^0 \to \rho^-\ell^+ \nu)/\Gamma (B^0
\to \pi^- \ell^+ \nu)$, are disfavoured by the CLEO data. The disfavoured
models are also those which introduce a larger theoretical dispersion in
the interpretation of the inclusive $B \to X_u \ell \nu_\ell$ and exclusive
decay data in terms of the ratio $\vert V_{ub}/V_{cb}\vert$. Excluding them
from further consideration, measurements in both the inclusive and
exclusive modes are compatible with
\beq
\left\vert \frac{V_{ub}}{V_{cb}} \right\vert = 0.08\pm 0.02~.
\label{vubvcbn}
\eeq
This gives
\beq
\sqrt{\rho^2 + \eta^2} = 0.36 \pm 0.08~.
\eeq
With the measurements of the form factors in semileptonic decays $B \to
(\pi,\rho,\omega) \ell \nu_\ell$, one should be able to further constrain
the models, thereby reducing the present theoretical uncertainty on this
quantity.

The experimental value of $\abseps$ is \cite{PDG}
\beq
\abseps = (2.26\pm 0.02)\times 10^{-3}~.
\eeq
Theoretically, $\abseps$ is essentially proportional to the imaginary part
of the box diagram for \kkbar\ mixing and is given by \cite{Burasetal}
\begin{eqnarray}
\abseps &=& \frac{G_F^2f_K^2M_KM_W^2}{6\sqrt{2}\pi^2\Delta M_K}
\hat{B}_K\left(A^2\lambda^6\eta\right)
\bigl(y_c\left\{\hat{\eta}_{ct}f_3(y_c,y_t)-\hat{\eta}_{cc}\right\}
 \nonumber \\
&~& ~~~~~~~~~~~~~~+ ~\hat{\eta}_{tt}y_tf_2(y_t)A^2\lambda^4(1-\rho)\bigr),
\label{eps}
\end{eqnarray}
where $y_i\equiv m_i^2/M_W^2$, and the functions $f_2$ and $f_3$ can be
found in Ref.~\cite{AL94}. Here, the $\hat{\eta}_i$ are QCD correction
factors, of which $\hat{\eta}_{cc}$ \cite{HN94} and $\hat{\eta}_{tt}$
\cite{etaB} were calculated some time ago to next-to-leading order, and
$\hat{\eta}_{ct}$ was known only to leading order \cite{Burastop,Flynn}.
Recently, this last renormalization constant was also calculated to
next-to-leading order \cite{HN95}. We use the following values for the
renormalization-scale-invariant coefficients: $\hat{\eta}_{cc}\simeq 1.32
$, $\hat{\eta}_{tt}\simeq 0.57$, $\hat{\eta}_{ct}\simeq 0.47 $, calculated
for $\hat{m}_c= 1.3$ GeV and  the NLO QCD parameter
$\Lambda_{\overline{MS}}=310$ MeV in Ref.~\cite{HN95}.

The final parameter in the expression for $\abseps$ is the
renormalization-scale independent parameter $\hat{B}_K$, which represents
our ignorance of the hadronic matrix element $\langle K^0 \vert
{({\overline{d}}\gamma^\mu (1-\gamma_5)s)}^2 \vert
{\overline{K^0}}\rangle$. The evaluation of this matrix element has been
the subject of much work. The earlier results are summarized in
Ref.~\cite{AL92}.

In our first set of fits, we consider specific values in the range 0.4 to
1.0 for $\hat{B}_K$. As we shall see, for $\hat{B}_K = 0.4$ a very poor fit
to the data is obtained, so that such small values are quite disfavoured.
In Fit 2, we assign a central value plus an error to $\hat{B}_K$. As in our
previous analysis \cite{AL94}, we consider two ranges for $\hat{B}_K$:
\beq
\hat{B}_K = 0.8 \pm 0.2 ~,
\label{BKrange1}
\eeq
which reflects the estimates of this quantity in lattice QCD
\cite{Shigemitsu,bklattice}, or
\beq
\hat{B}_K = 0.6 \pm 0.2 ~,
\label{BKrange2}
\eeq
which overlaps with the values suggested by chiral perturbation theory
\cite{Pich94}. As we will see, there is not an enormous difference in the
results for the two ranges.

We now turn to \bdbdbar\ mixing. The present world average of $\xd\equiv
\Delta M_d/\Gamma_d$, which is a measure of this mixing, is \cite{Wells}
\beq
\xd = 0.73 \pm 0.04~,
\label{xdvalue}
\eeq
which is based on time-integrated measurements which directly measure
$\xd$, and on time-dependent measurements which measure the mass difference
$\Delta M_d$ directly. This is then converted to $\xd$ using the $B_d^0$
lifetime, which is known very precisely $(\tau(B_d)=1.57 \pm 0.05~ps$).
{}From a theoretical point of view it is better to use the mass difference
$\Delta M_d$, as it liberates one from the errors on the lifetime
measurement. In fact, the present precision on $\Delta M_d$, pioneered by
time-dependent techniques at LEP, is quite competitive with the precision
on $\xd$. The LEP average for $\Delta M_d$ has been combined with that
derived from time-integrated measurements yielding the present world
average \cite{Wells}
\beq
\Delta M_d = 0.465 \pm 0.024~(ps)^{-1} ~.
\label{deltamd}
\eeq
We shall use this number instead of $\xd$, which has been the usual
practice to date \cite{Burastop,AL92,Pich94,BLO94}.

\begin{table}
\hfil
\vbox{\offinterlineskip
\halign{&\vrule#&
 \strut\quad#\hfil\quad\cr
\noalign{\hrule}
height2pt&\omit&&\omit&\cr
& Parameter && Value & \cr
height2pt&\omit&&\omit&\cr
\noalign{\hrule}
height2pt&\omit&&\omit&\cr
& $\lambda$ && $0.2205$ & \cr
& $\vert V_{cb} \vert $ && $0.0388 \pm 0.0036$ & \cr
& $\vert V_{ub} / V_{cb} \vert$ && $0.08 \pm 0.02$ & \cr
& $\abseps$ && $(2.26 \pm 0.02) \times 10^{-3}$ & \cr
& $\Delta M_d$ && $(0.465 \pm 0.024)~(ps)^{-1}$ & \cr
& $\tau(B_d)$ && $(1.57 \pm 0.05)~(ps)$ & \cr
& $\overline{\mt}(\mt(pole))$ && $(170 \pm 11)$ GeV & \cr
& $\hat{\eta}_B$ && $0.55$ & \cr
& $\hat{\eta}_{cc} $ && $1.32$ & \cr
& $\hat{\eta}_{ct} $ && $0.47$ & \cr
& $\hat{\eta}_{tt} $ && $0.57$ & \cr
& $\hat{B}_K$ && $0.8 \pm 0.2$ & \cr
& $\hat{B}_B$ && $1.0 \pm 0.2$ & \cr
& $\fbd$ && $180 \pm 50$ MeV & \cr
height2pt&\omit&&\omit&\cr
\noalign{\hrule}}}
\caption{Parameters used in the CKM fits. Values of the hadronic quantities
$\fbd$, $\hat{B}_{B_d}$ and $\hat{B}_K$ shown are motivated by the lattice
QCD results. In Fit 1, specific values of these hadronic quantities are
chosen, while in Fit 2, they are allowed to vary over the given ranges. (In
Fit 2, for comparison we also consider the range $\hat{B}_K = 0.6 \pm 0.2$,
which is motivated by chiral perturbation theory and QCD sum rules.)}
\label{tabfit}
\end{table}

The mass difference $\Delta M_d$ is calculated from the \bdbdbar\ box
diagram. Unlike the kaon system, where the contributions of both the $c$-
and the $t$-quarks in the loop were important, this diagram is dominated by
$t$-quark exchange:
\beq
\label{bdmixing}
\Delta M_d = \frac{G_F^2}{6\pi^2}M_W^2M_B\left(\fbb\right)\hat{\eta}_B y_t
f_2(y_t) \vert V_{td}^*V_{tb}\vert^2~, \label{xd}
\eeq
where, using Eq.~\ref{CKM}, $\vert V_{td}^*V_{tb}\vert^2=
A^2\lambda^{6}\left[\left(1-\rho\right)^2+\eta^2\right]$. Here,
$\hat{\eta}_B$ is the QCD correction. In Ref.~\cite{etaB}, this correction
is analyzed including the effects of a heavy $t$-quark. It is found that
$\hat{\eta}_B$ depends sensitively on the definition of the $t$-quark mass,
and that, strictly speaking, only the product $\hat{\eta}_B(y_t)f_2(y_t)$
is free of this dependence. In the fits presented here we use the value
$\hat{\eta}_B=0.55$, calculated in the $\overline{MS}$ scheme, following
Ref.~\cite{etaB}. Consistency requires that the top quark mass be rescaled
from its pole (mass) value of $\mt =180 \pm 11$ GeV to the value
$\overline{\mt}(\mt(pole))$ in the $\overline{MS}$ scheme, which is
typically about 10 GeV smaller \cite{mtmsbar}.

For the $B$ system, the hadronic uncertainty is given by $\fbb$, analogous
to $\hat{B}_K$ in the kaon system, except that in this case, also $\fbd$ is
not measured. In our fits, we will take ranges for $\fbb$ and
$\hat{B}_{B_d}$ which are compatible with results from both lattice-QCD and
QCD sum rules \cite{Shigemitsu,Narison,Sommer94}:
\begin{eqnarray}
\fbd &=& 180 \pm 50 ~\mbox{MeV}~, \nonumber \\
\hat{B}_{B_d} &=& 1.0 \pm 0.2 ~.
\label{FBrange}
\end{eqnarray}
In Table \ref{tabfit}, we summarize all input quantities to our fits, of
which seven quantities ($\absvcb$, $\vert V_{ub}/V_{cb} \vert$, $\delmd$,
$\tau(B_d)$, ${\overline{\mt}}$, $\hat{\eta}_{cc}$, $\hat{\eta}_{ct}$)
 have changed compared to
their values used in our previous fit \cite{AL94}.


\section{The Unitarity Triangle}

\begin{figure}
\centerline{\psfig{figure=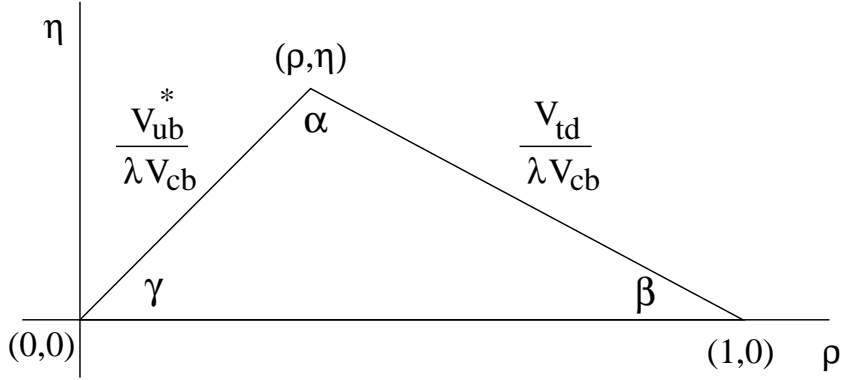,height=5.0cm,angle=90}}
\caption{The unitarity triangle. The angles $\alpha$, $\beta$ and $\gamma$
can be measured via CP violation in the $B$ system.}
\label{triangle}
\end{figure}

The allowed region in $\rho$-$\eta$ space can be displayed quite elegantly
using the so-called unitarity triangle. The unitarity of the CKM matrix
leads to the following relation:
\beq
V_{ud} V_{ub}^* + V_{cd} V_{cb}^* + V_{td} V_{tb}^* = 0~.
\eeq
Using the form of the CKM matrix in Eq.~\ref{CKM}, this can be recast as
\beq
\frac{V_{ub}^*}{\lambda V_{cb}} + \frac{V_{td}}{\lambda V_{cb}} = 1~,
\eeq
which is a triangle relation in the complex plane (i.e.\ $\rho$-$\eta$
space), illustrated in Fig.~\ref{triangle}. Thus, allowed values of $\rho$
and $\eta$ translate into allowed shapes of the unitarity triangle.

In order to find the allowed unitarity triangles, the computer program
MINUIT is used to fit the CKM parameters $A$, $\rho$ and $\eta$ to the
experimental values of $\absvcb$, $\vert V_{ub}/V_{cb}\vert$, $\abseps$ and
$\xd$. Since $\lambda$ is very well measured, we have fixed it to its
central value given above. As discussed in the introduction, we present
here two types of fits:
\begin{itemize}
\item
Fit 1: the ``experimental fit.'' Here, only the experimentally measured
numbers are used as inputs to the fit with Gaussian errors; the coupling
constants $f_{B_d} \sqrt{\hat{B}_{B_d}}$ and $\hat{B}_K$ are given fixed
values.
\item
Fit 2: the ``combined fit.'' Here, both the experimental and theoretical
numbers are used as inputs assuming Gaussian errors for the theoretical
quantities.
\end{itemize}

We first discuss the ``experimental fit" (Fit 1). The goal here is to
restrict the allowed range of the parameters ($\rho,\eta)$ for given values
of the coupling constants $f_{B_d} \sqrt{\hat{B}_{B_d}}$ and $\hat{B}_K$.
For each value of $\hat{B}_K$ and $f_{B_d}\sqrt{\hat{B}_{B_d}}$, the CKM
parameters $A$, $\rho$ and $\eta$ are fit to the experimental numbers given
in Table \ref{tabfit} and the $\chi^2$ is calculated.

\begin{figure}
\vskip -2.4truein
\centerline{\epsfxsize 7.0 truein \epsfbox {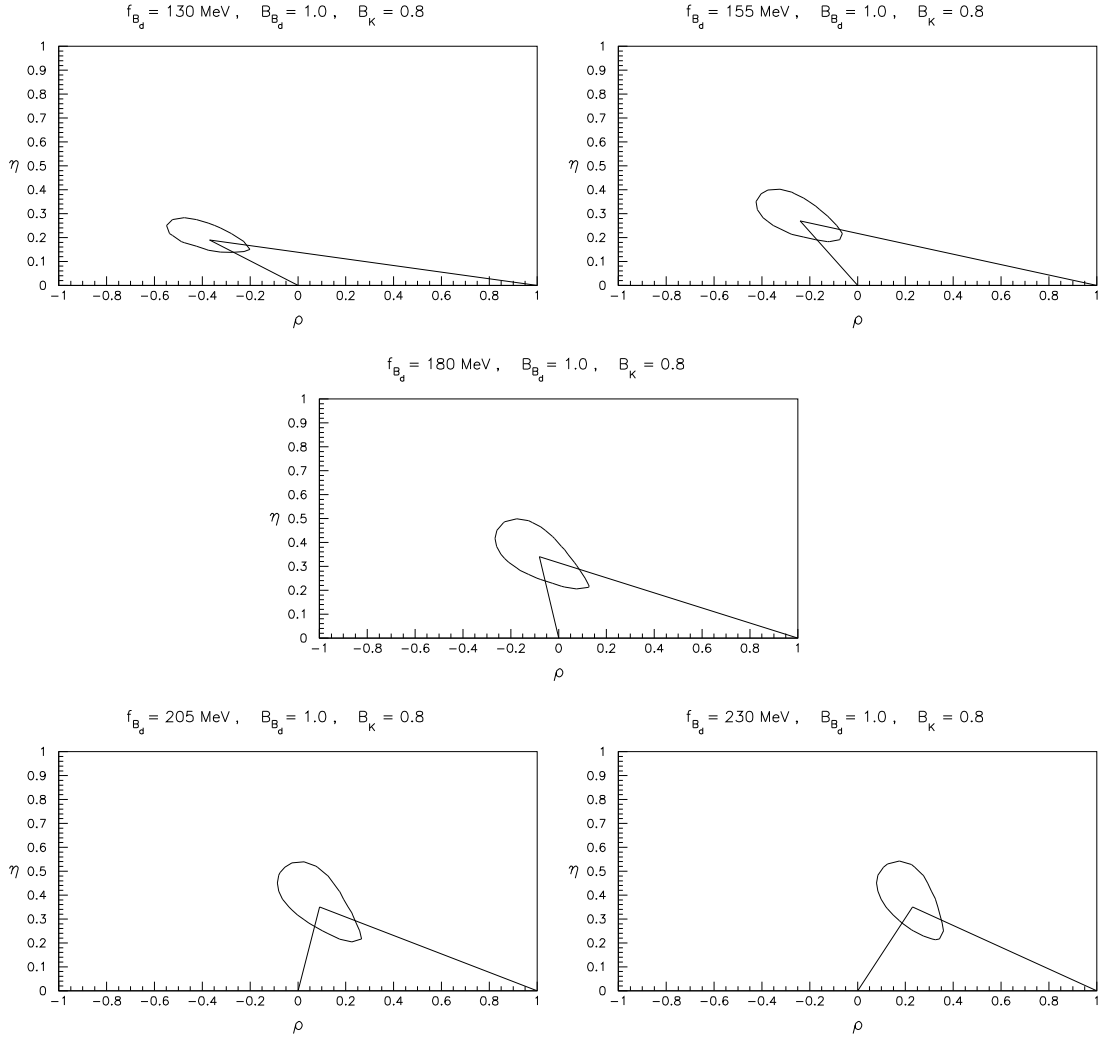}}
\vskip -1.0truein
\caption{Allowed region in $\rho$-$\eta$ space, from a fit to the
experimental values given in Table \protect{\ref{tabfit}}. We have fixed
$\hat{B}_K=0.8$ and vary the coupling constant product
$\fbd\protect\sqrt{\hat{B}_{B_d}}$ as indicated on the figures. The solid
line represents the region with $\chi^2=\chi_{min}^2+6$ corresponding to
the 95\% C.L.\ region. The triangles show the best fit.}
\label{rhoeta1}
\end{figure}

First, we fix $\hat{B}_K = 0.8$, and vary $f_{B_d}\sqrt{\hat{B}_{B_d}}$ in
the range 130 MeV to 230 MeV. The fits are presented as an allowed region
in $\rho$-$\eta$ space at 95\% C.L. ($\chi^2 = \chi^2_{min} + 6.0$). The
results are shown in Fig.~\ref{rhoeta1}. As we pass from
Fig.~\ref{rhoeta1}(a) to Fig.~\ref{rhoeta1}(e), the unitarity triangles
represented by these graphs become more and more obtuse. Even more striking
than this, however, is the fact that the range of possibilities for these
triangles is quite large. There are two things to be learned from this.
First, our knowledge of the unitarity triangle is at present rather poor.
This will be seen even more clearly when we present the results of Fit 2.
Second, unless our knowledge of hadronic matrix elements improves
considerably, measurements of $\abseps$ and $x_d$, no matter how precise,
will not help much in further constraining the unitarity triangle. This is
why measurements of CP-violating rate asymmetries in the $B$ system are so
important \cite{BCPasym,AKL94}. Being largely independent of theoretical
uncertainties, they will allow us to accurately pin down the unitarity
triangle. With this knowledge, we could deduce the correct values of
$\hat{B}_K$ and $f_{B_d}\sqrt{\hat{B}_{B_d}}$, and thus rule out or confirm
different theoretical approaches to calculating these hadronic quantities.

Despite the large allowed region in the $\rho$-$\eta$ plane, certain values
of $\hat{B}_K$ and $f_{B_d}\sqrt{\hat{B}_{B_d}}$ are disfavoured since they
do not provide a good fit to the data. For example, fixing $\hat{B}_K=1.0$,
we can use the fitting program to provide the minimum $\chi^2$ for various
values of $f_{B_d}\sqrt{\hat{B}_{B_d}}$. The results are shown in Table
\ref{tabbk1}, along with the best fit values of $(\rho,\eta)$. Since we
have two variables ($\rho$ and $\eta$), we use $\chi^2_{min}<2.0$ as our
``good fit" criterion, and we see that $f_{B_d} \sqrt{\hat{B}_{B_d}} < 130$
MeV and $f_{B_d} \sqrt{\hat{B}_{B_d}} > 270$ MeV give poor fits to the
existing data. Note also that the $\chi^2$ distribution has two minima, at
around $f_{B_d} \sqrt{\hat{B}_{B_d}} = 150$ and 230 MeV. We do not consider
this terribly significant, since the surrounding values of $f_{B_d}
\sqrt{\hat{B}_{B_d}}$ also yield good fits to the data. The very small
values of $\chi^2_{min}$ depend sensitively on the central values of the
various experimental quantities -- if these values move around a little
bit, the values of $f_{B_d} \sqrt{\hat{B}_{B_d}}$ which give the minimum
$\chi^2$ values will move around as well. In Tables \ref{tabbk8} and
\ref{tabbk6}, we present similar analyses, but for $\hat{B}_K=0.8$ and
$0.6$, respectively. From these tables we see that the lower limit on
$f_{B_d} \sqrt{\hat{B}_{B_d}}$ remains fairly constant, at around 130 MeV,
but the upper limit depends quite strongly on the value of $\hat{B}_K$
chosen. Specifically, for $\hat{B}_K =0.8$ and $0.6$, the maximum allowed
value of $f_{B_d} \sqrt{\hat{B}_{B_d}}$ is about 240 and 210 MeV,
respectively.

In Table \ref{tabbk4}, we present the $\chi^2$ values as a function of
$f_{B_d} \sqrt{\hat{B}_{B_d}}$ for $\hat{B}_K=0.4$, which is not favoured
by lattice calculations or QCD sum rules. What is striking is that, over
the entire range of $f_{B_d} \sqrt{\hat{B}_{B_d}}$, the minimum $\chi^2$ is
always greater than 2. This indicates that the data strongly disfavour
$\hat{B}_K \leq 0.4$ solutions.

\begin{table}
\hfil
\vbox{\offinterlineskip
\halign{&\vrule#&
 \strut\quad#\hfil\quad\cr
\noalign{\hrule}
height2pt&\omit&&\omit&&\omit&\cr
& $\fbd\sqrt{\hat{B}_{B_d}}$ (MeV) && $(\rho,\eta)$ && $\chi^2_{min}$ & \cr
height2pt&\omit&&\omit&&\omit&\cr
\noalign{\hrule}
height2pt&\omit&&\omit&&\omit&\cr
& $120$ && $(-0.43,~0.13)$ && $2.89$ & \cr
& $130$ && $(-0.38,~0.16)$ && $1.16$ & \cr
& $140$ && $(-0.34,~0.18)$ && $0.3$ & \cr
& $150$ && $(-0.30,~0.21)$ && $9.2 \times 10^{-3}$ & \cr
& $160$ && $(-0.25,~0.24)$ && $0.07$ & \cr
& $170$ && $(-0.20,~0.27)$ && $0.29$ & \cr
& $180$ && $(-0.14,~0.29)$ && $0.52$ & \cr
& $190$ && $(-0.07,~0.31)$ && $0.64$ & \cr
& $200$ && $(0.0,~0.31)$ && $0.59$ & \cr
& $210$ && $(0.07,~0.31)$ && $0.38$ & \cr
& $220$ && $(0.13,~0.31)$ && $0.13$ & \cr
& $230$ && $(0.19,~0.3)$ && $3.8 \times 10^{-3}$ & \cr
& $240$ && $(0.23,~0.31)$ && $0.07$ & \cr
& $250$ && $(0.27,~0.31)$ && $0.36$ & \cr
& $260$ && $(0.31,~0.31)$ && $0.87$ & \cr
& $270$ && $(0.34,~0.31)$ && $1.59$ & \cr
& $280$ && $(0.38,~0.32)$ && $2.51$ & \cr
height2pt&\omit&&\omit&&\omit&\cr
\noalign{\hrule}}}
\caption{The ``best values'' of the CKM parameters $(\rho,\eta)$ as a
function of the coupling constant $\fbd\protect\sqrt{\hat{B}_{B_d}}$,
obtained by a minimum $\chi^2$ fit to the experimental data, including the
renormalized value of $m_t=170 \pm 11$ GeV. We fix $\hat{B}_K=1.0$. The
resulting minimum $\chi^2$ values from the MINUIT fits are also given.}
\label{tabbk1}
\end{table}

\begin{table}
\hfil
\vbox{\offinterlineskip
\halign{&\vrule#&
 \strut\quad#\hfil\quad\cr
\noalign{\hrule}
height2pt&\omit&&\omit&&\omit&\cr
& $\fbd\sqrt{\hat{B}_{B_d}}$ (MeV) && $(\rho,\eta)$ && $\chi^2_{min}$ & \cr
height2pt&\omit&&\omit&&\omit&\cr
\noalign{\hrule}
height2pt&\omit&&\omit&&\omit&\cr
& $120$ && $(-0.42,~0.16)$ && $3.04$ & \cr
& $130$ && $(-0.37,~0.19)$ && $1.32$ & \cr
& $140$ && $(-0.32,~0.23)$ && $0.43$ & \cr
& $150$ && $(-0.27,~0.26)$ && $0.07$ & \cr
& $160$ && $(-0.22,~0.29)$ && $1.4 \times 10^{-3}$ & \cr
& $170$ && $(-0.15,~0.32)$ && $0.05$ & \cr
& $180$ && $(-0.08,~0.34)$ && $0.09$ & \cr
& $190$ && $(-0.01,~0.35)$ && $0.06$ & \cr
& $200$ && $(0.06,~0.35)$ && $0.01$ & \cr
& $210$ && $(0.13,~0.35)$ && $0.02$ & \cr
& $220$ && $(0.18,~0.35)$ && $0.2$ & \cr
& $230$ && $(0.23,~0.35)$ && $0.61$ & \cr
& $240$ && $(0.28,~0.35)$ && $1.29$ & \cr
& $250$ && $(0.32,~0.35)$ && $2.22$ & \cr
height2pt&\omit&&\omit&&\omit&\cr
\noalign{\hrule}}}
\caption{The ``best values'' of the CKM parameters $(\rho,\eta)$ as a
function of the coupling constant $\fbd\protect\sqrt{\hat{B}_{B_d}}$,
obtained by a minimum $\chi^2$ fit to the experimental data, including the
renormalized value of $m_t=170 \pm 11$ GeV. We fix $\hat{B}_K=0.8$. The
resulting minimum $\chi^2$ values from the MINUIT fits are also given.}
\label{tabbk8}
\end{table}

\begin{table}
\hfil
\vbox{\offinterlineskip
\halign{&\vrule#&
 \strut\quad#\hfil\quad\cr
\noalign{\hrule}
height2pt&\omit&&\omit&&\omit&\cr
& $\fbd\sqrt{\hat{B}_{B_d}}$ (MeV) && $(\rho,\eta)$ && $\chi^2_{min}$ & \cr
height2pt&\omit&&\omit&&\omit&\cr
\noalign{\hrule}
height2pt&\omit&&\omit&&\omit&\cr
& $120$ && $(-0.4,~0.21)$ && $3.39$ & \cr
& $130$ && $(-0.35,~0.25)$ && $1.7$ & \cr
& $140$ && $(-0.29,~0.29)$ && $0.78$ & \cr
& $150$ && $(-0.22,~0.33)$ && $0.35$ & \cr
& $160$ && $(-0.15,~0.36)$ && $0.18$ & \cr
& $170$ && $(-0.07,~0.38)$ && $0.16$ & \cr
& $180$ && $(0.01,~0.39)$ && $0.24$ & \cr
& $190$ && $(0.08,~0.4)$ && $0.48$ & \cr
& $200$ && $(0.15,~0.4)$ && $0.96$ & \cr
& $210$ && $(0.21,~0.4)$ && $1.73$ & \cr
& $220$ && $(0.26,~0.4)$ && $2.85$ & \cr
height2pt&\omit&&\omit&&\omit&\cr
\noalign{\hrule}}}
\caption{The ``best values'' of the CKM parameters $(\rho,\eta)$ as a
function of the coupling constant $\fbd\protect\sqrt{\hat{B}_{B_d}}$,
obtained by a minimum $\chi^2$ fit to the experimental data, including the
renormalized value of $m_t=170 \pm 11$ GeV. We fix $\hat{B}_K=0.6$. The
resulting minimum $\chi^2$ values from the MINUIT fits are also given.}
\label{tabbk6}
\end{table}

\begin{table}
\hfil
\vbox{\offinterlineskip
\halign{&\vrule#&
 \strut\quad#\hfil\quad\cr
\noalign{\hrule}
height2pt&\omit&&\omit&&\omit&\cr
& $\fbd\sqrt{\hat{B}_{B_d}}$ (MeV) && $(\rho,\eta)$ && $\chi^2_{min}$ & \cr
height2pt&\omit&&\omit&&\omit&\cr
\noalign{\hrule}
height2pt&\omit&&\omit&&\omit&\cr
& $130$ && $(-0.28,~0.35)$ && $2.95$ & \cr
& $140$ && $(-0.2,~0.39)$ && $2.22$ & \cr
& $150$ && $(-0.11,~0.43)$ && $2.04$ & \cr
& $160$ && $(-0.02,~0.45)$ && $2.32$ & \cr
& $170$ && $(0.06,~0.46)$ && $3.07$ & \cr
height2pt&\omit&&\omit&&\omit&\cr
\noalign{\hrule}}}
\caption{The ``best values'' of the CKM parameters $(\rho,\eta)$ as a
function of the coupling constant $\fbd\protect\sqrt{\hat{B}_{B_d}}$,
obtained by a minimum $\chi^2$ fit to the experimental data, including the
renormalized value of $m_t=170 \pm 11$ GeV. We fix $\hat{B}_K=0.4$. The
resulting minimum $\chi^2$ values from the MINUIT fits are also given.}
\label{tabbk4}
\end{table}

\begin{figure}
\vskip -1.0truein
\centerline{\epsfxsize 3.5 truein \epsfbox {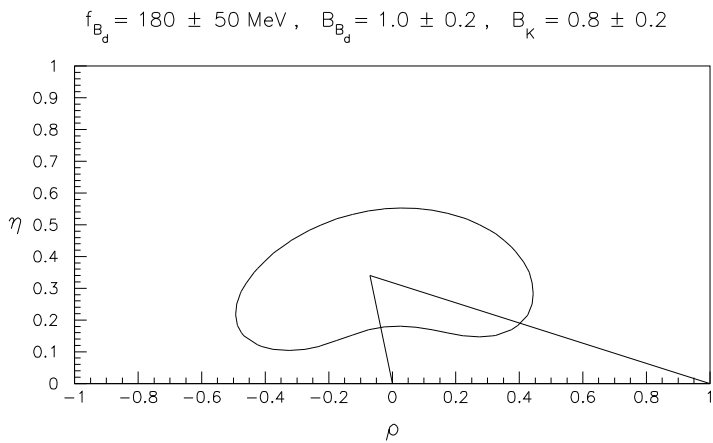}}
\vskip -1.4truein
\caption{Allowed region in $\rho$-$\eta$ space, from a simultaneous fit to
both the experimental and theoretical quantities given in Table
\protect{\ref{tabfit}}. The theoretical errors are treated as Gaussian for
this fit. The solid line represents the region with $\chi^2=\chi_{min}^2+6$
corresponding to the 95\% C.L.\ region. The triangle shows the best fit.}
\label{rhoeta2}
\end{figure}

We now discuss the ``combined fit" (Fit 2). Since the coupling constants
are not known and the best we have are estimates given in the ranges in
Eqs.~(\ref{BKrange1}) and (\ref{FBrange}), a reasonable profile of the
unitarity triangle at present can be obtained by letting the coupling
constants vary in these ranges. The resulting CKM triangle region is shown
in Fig.~\ref{rhoeta2}. As is clear from this figure, the allowed region is
rather large at present. The
preferred values obtained from the ``combined fit" are
\beq
(\rho,\eta) = (-0.07,0.34) ~~~(\mbox{with}~\chi^2 = 6.6\times 10^{-2})~.
\eeq

\begin{figure}
\vskip -1.0truein
\centerline{\epsfxsize 3.5 truein \epsfbox {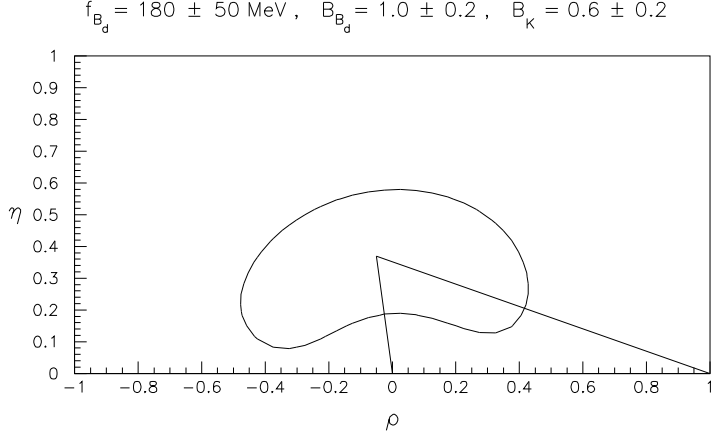}}
\vskip -1.4truein
\caption{Allowed region in $\rho$-$\eta$ space, from a simultaneous fit to
both the experimental and theoretical quantities given in Table
\protect{\ref{tabfit}}, except that we take $\hat{B}_K = 0.6 \pm 0.2$. The
theoretical errors are treated as Gaussian for this fit. The solid line
represents the region with $\chi^2=\chi_{min}^2+6$ corresponding to the
95\% C.L.\ region. The triangle shows the best fit.}
\label{rhoeta3}
\end{figure}

For comparison, we also show the allowed region in the $(\rho,\eta)$ plane
for the case in which $\hat{B}_K = 0.6 \pm 0.2$ [Eq.~(\ref{BKrange2})],
which is more favoured by chiral perturbation theory and QCD sum rules. The
CKM triangle region is shown in Fig.~\ref{rhoeta3}. Clearly, there is not
much difference between this figure and Fig.~\ref{rhoeta2}. The preferred
values obtained from this fit are
\beq
(\rho,\eta) = (-0.05,0.37) ~~~(\mbox{with}~\chi^2 = 0.1)~.
\eeq

\section{$\xs$ and the Unitarity Triangle}

Mixing in the \bsbsbar\ system is quite similar to that in the \bdbdbar\
system. The \bsbsbar\ box diagram is again dominated by $t$-quark exchange,
and the mass difference between the mass eigenstates $\delms$ is given by a
formula analogous to that of Eq.~(\ref{xd}):
\beq
\delms = \frac{G_F^2}{6\pi^2}M_W^2M_{B_s}\left(\fbbs\right)
\hat{\eta}_{B_s} y_t f_2(y_t) \vert V_{ts}^*V_{tb}\vert^2~.
\label{xs}
\eeq
Using the fact that $\vert V_{cb}\vert=\vert V_{ts}\vert$ (Eq.~\ref{CKM}),
it is clear that one of the sides of the unitarity triangle, $\vert
V_{td}/\lambda V_{cb}\vert$, can be obtained from the ratio of $\delmd$ and
$\delms$,
\beq
\frac{\delms}{\delmd} =
 \frac{\hat{\eta}_{B_s}M_{B_s}\left(\fbbs\right)}
{\hat{\eta}_{B_d}M_{B_d}\left(\fbb\right)}
\left\vert \frac{V_{ts}}{V_{td}} \right\vert^2.
\label{xratio}
\eeq
All dependence on the $t$-quark mass drops out, leaving the square of the
ratio of CKM matrix elements, multiplied by a factor which reflects
$SU(3)_{\rm flavour}$ breaking effects. The only real uncertainty in this
factor is the ratio of hadronic matrix elements. Whether or not $\xs$ can
be used to help constrain the unitarity triangle will depend crucially on
the theoretical status of the ratio $\fbbs/\fbb$. In what follows, we will
take $\xi_s \equiv (f_{B_s} \sqrt{\hat{B}_{B_s}}) / (f_{B_d}
\sqrt{\hat{B}_{B_d}}) = (1.16 \pm 0.1)$, consistent with both lattice-QCD
\cite{Shigemitsu} and QCD sum rules \cite{Narison}. (The SU(3)-breaking
factor in $\delms/\delmd$ is $\xi_s^2$.)

The mass and lifetime of the $B_s$ meson have now been measured at LEP and
Tevatron and their present values are $M_{B_s}=5370.0 \pm 2.0$ MeV and
$\tau(B_s)= 1.58 \pm 0.10 ~ps$ \cite{komamiya95}. We expect
the QCD correction factor $\hat{\eta}_{B_s}$ to be equal to its $B_d$
counterpart, i.e.\ $\hat{\eta}_{B_s} =0.55$. The main uncertainty in $\xs$
(or, equivalently, $\delms$) is now $\fbbs$. Using the determination of $A$
given previously, $\tau_{B_s}= 1.58 \pm 0.10~(ps)$ and $\overline{\mt}=171
\pm 11$ GeV, we obtain
\begin{eqnarray}
\delms &=& \left(13.1 \pm 2.8\right)\frac{\fbbs}{(230~\mbox{MeV})^2}
{}~(ps)^{-1}~, \nonumber \\
\xs &=& \left(20.7 \pm 4.5\right)\frac{\fbbs}{(230~\mbox{MeV})^2}~.
\end{eqnarray}
The choice $f_{B_s}\sqrt{\hat{B}_{B_s}}= 230$ MeV corresponds to the
central value given by the lattice-QCD estimates, and with this our fits
give $\xs \simeq 20$ as the preferred value in the SM. Allowing the
coefficient to vary by $\pm 2\sigma$, and
taking the central value for
$f_{B_s}\sqrt{\hat{B}_{B_s}}$, this gives
\begin{eqnarray}
11.7 &\leq & \xs \leq 29.7~, \nonumber\\
7.5 ~(ps)^{-1} &\leq & \delms \leq 18.7 ~(ps)^{-1}~.
 \label{bestxs}
\end{eqnarray}
It is difficult to ascribe a confidence level to this range due to the
dependence on the unknown coupling constant factor. All one can say is that
the standard model predicts large values for $\xs$, most of which are above
the present experimental limit $\xs > 8.8$ (equivalently $\delms > 6.1
{}~(ps)^{-1}$) \cite{ALEPHxs}.

\begin{figure}
\vskip -1.0truein
\centerline{\epsfxsize 3.5 truein \epsfbox {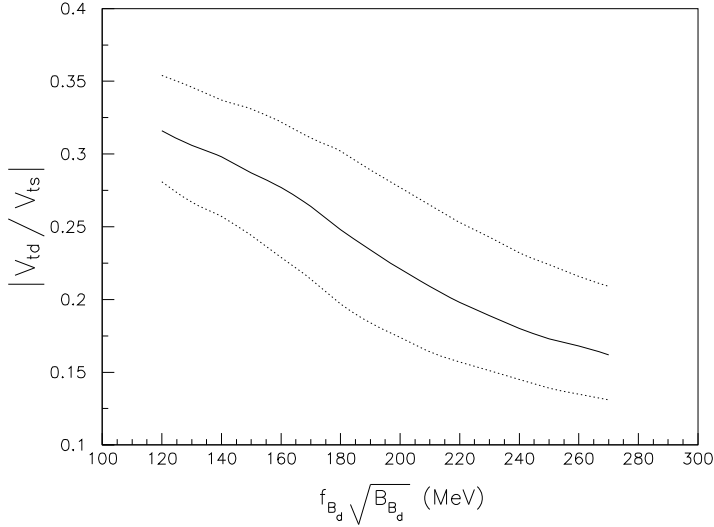}}
\vskip -1.0truein
\caption{Allowed values of the CKM matrix element ratio $\vert
V_{td}/V_{ts} \vert$ as a function of the coupling constant product
$f_{B_d}\protect\sqrt{\hat{B}_{B_d}}$, for $\hat{B}_K=0.8\pm 0.2$. The solid
line corresponds to the best fit values and the dotted curves correspond to
the maximum and minimum allowed values at 95 \% C.L.}
\label{vtdts}
\end{figure}

An alternative estimate of $\delms$ (or $\xs$) can also be obtained by
using the relation in Eq.~(\ref{xratio}). Two quantities are required.
First, we need the CKM ratio $\vert V_{ts}/V_{td} \vert$. In
Fig.~\ref{vtdts} we show the allowed values (at 95\% C.L.) of the inverse
of this ratio as a function of $\fbd\sqrt{\hat{B}_{B_d}}$, for
$\hat{B}_K=0.8\pm 0.2$. From this one gets
\beq
2.8 \leq \left\vert {V_{ts} \over V_{td}} \right\vert \leq 7.6~.
\eeq
The second ingredient is the SU(3)-breaking factor which we take to be
$\xi_s = 1.16 \pm 0.1$, or $1.1 \le \xi_s^2 \le 1.6$.
 The result of the CKM fit can
 therefore be expressed as a $95\%$ C.L. range:
\beq
 10.5 \left(\frac{\xi_s}{1.16}\right)^2
 ~\leq ~\frac{\delms}{\delmd} ~\leq ~
 77.7 \left(\frac{\xi_s}{1.16}\right)^2 ~.
\eeq
Again, it is difficult to assign a true confidence level to $\delms/\delmd$
due to the dependence on $\xi_s$. The large allowed range reflects our poor
knowledge of the matrix element ratio $\vert V_{ts}/V_{td} \vert$, which
shows that this method is not particularly advantageous at present for the
determination of the range for $\delms$.

The ALEPH lower bound $\delms > 6.1~(ps)^{-1}$ (95\% C.L.) \cite{ALEPHxs}
and the present world average $\delmd = (0.465 \pm 0.024)~(ps)^{-1}$ can
be used to put a bound on the ratio $\delms/\delmd$. The lower limit on
$\delms$ is correlated with the value of $f_s$, the fraction of $b$ quark
fragmenting into $B_s$ meson, as shown in the ALEPH analysis
\cite{ALEPHxs}. The value obtained from the measurement  of the quantity
$f_s BR(B_s \to D_s \ell \nu_\ell)$ is $f_s=(12 \pm 3) \%$.  The
time-integrated mixing ratios $\bar{\chi}$ and $\chi_d$, assuming maximal
mixing in the $B_s$-$\overline{B_s}$ system $\chi_s =0.5$, give $f_s=(9\pm
2)\%$. The weighted average of these numbers is $f_s=(10 \pm 2)\%$
\cite{Wells}. With $f_s=10 \%$, one gets $\delms > 5.6~(ps)^{-1}$ at 95\%
C.L., yielding $\delms/\delmd > 11.3$ at 95\% C.L. Assuming, however,
$f_s=12\%$ gives $\delms > 6.1~(ps)^{-1}$, yielding $\delms/\delmd >
12.3$ at 95\% C.L. We will use this latter number.

\begin{figure}
\vskip -1.0truein
\centerline{\epsfxsize 3.5 truein \epsfbox {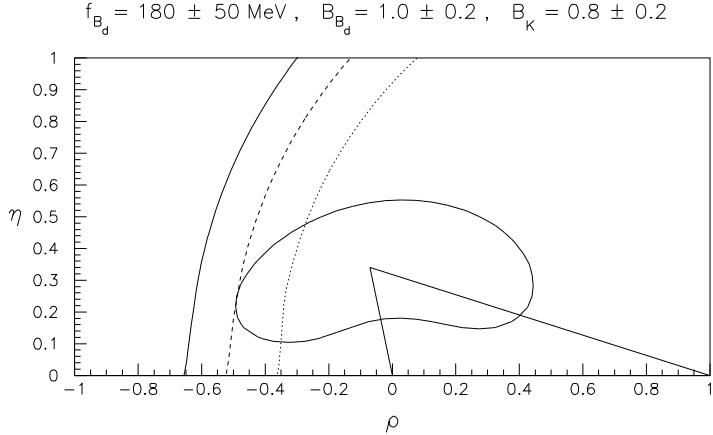}}
\vskip -1.4truein
\caption{Further constraints in $\rho$-$\eta$ space from the ALEPH bound
on $\delms$. The bounds are presented for 3 choices of the SU(3)-breaking
parameter: $\xi_s^2 = 1.1$ (dotted line), $1.35$ (dashed line) and $1.6$
(solid line). In all cases, the region to the left of the curve is ruled
out.}
\label{xslimit}
\end{figure}

The 95\% confidence limit on $\delms/\delmd$ can be turned into a bound on
the CKM parameter space $(\rho,\eta)$ by choosing a value for the
SU(3)-breaking parameter $\xi_s^2$. We assume three representative values:
$\xi_s^2 = 1.1$, $1.35$ and $1.6$, and display the resulting constraints in
Fig.~\ref{xslimit}. From this graph we see that the ALEPH bound marginally
restricts the allowed $\rho$-$\eta$ region for small values of $\xi_s^2$,
but does not provide any useful bounds for larger values.

Summarizing the discussion on $\xs$, we note that the lattice-QCD-inspired
estimate $f_{B_s} \sqrt{\hat{B}_{B_s}} \simeq 230$ MeV and the CKM fit
predict that $\xs$ lies between 12 and 30, with a central value around 20.
The upper and lower bounds and the central value scale as
$(f_{B_s}\sqrt{\hat{B}_{B_s}}/230 ~\mbox{MeV})^2$. The present constraints
from the lower bound on $\xs$ on the CKM parameters are marginal but this
would change with improved data. In particular, one expects to reach a
sensitivity $\xs \simeq 15$ (or $\delms \simeq 10~ps^{-1})$ at LEP
combining all data and
 tagging techniques \cite{Wells},
which would be in the ball-park estimate for this
quantity in the SM presented here.
 Of course, an actual measurement of $\xs$
would be very helpful in further constraining the CKM parameter space.


\section{CP Violation in the $B$ System}

It is expected that the $B$ system will exhibit large CP-violating effects,
characterized by nonzero values of the angles $\alpha$, $\beta$ and
$\gamma$ in the unitarity triangle (Fig.~\ref{triangle}) \cite{BCPasym}.
The most promising method to measure CP violation is to look for an
asymmetry between $\Gamma(B^0\to f)$ and $\Gamma({\overline{B^0}}\to f)$,
where $f$ is a CP eigenstate. If only one weak amplitude contributes to the
decay, the CKM phases can be extracted cleanly (i.e.\ with no hadronic
uncertainties). Thus, $\sin 2\alpha$, $\sin 2\beta$ and $\sin 2\gamma$ can
in principle be measured in $\bdbarp \to \pi^+ \pi^-$, $\bdbarp\to J/\psi
K_S$ and $\bsbarp\to\rho K_S$, respectively.

Unfortunately, the situation is not that simple. In all of the above cases,
in addition to the tree contribution, there is an additional amplitude due
to penguin diagrams \cite{penguins}. In general, this will introduce some
hadronic uncertainty into an otherwise clean measurement of the CKM phases.
In the case of $\bdbarp\to J/\psi K_S$, the penguins do not cause any
problems, since the weak phase of the penguin is the same as that of the
tree contribution. Thus, the CP asymmetry in this decay still measures
$\sin 2\beta$.

\begin{figure}
\vskip -1.0truein
\centerline{\epsfxsize 3.5 truein \epsfbox {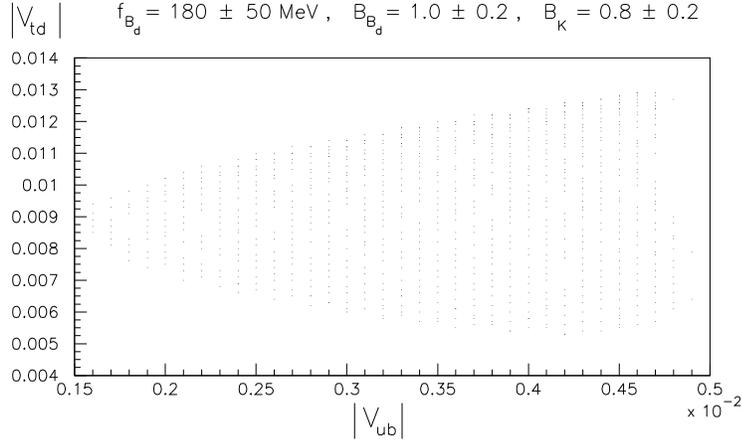}}
\vskip -1.4truein
\caption{Allowed region of the CKM matrix elements $|V_{td}|$ and
$|V_{ub}|$ resulting from the ``combined fit" of the data for the ranges
for $\fbd\protect\sqrt{\hat{B}_{B_d}} $ and $\hat{B}_K$ given in the text.}
\label{vtdvub}
\end{figure}

For $\bdbarp \to \pi^+ \pi^-$, however, although the penguin is expected to
be small with respect to the tree diagram, it will still introduce a
theoretical uncertainty into the extraction of $\alpha$. Fortunately, this
uncertainty can be removed by the use of isospin \cite{isospin}. The key
observation is that the $I=2$ component of the $B\to\pi\pi$ amplitude is
pure tree (i.e., it has no penguin contribution) and therefore has a
well-defined CKM phase. By measuring the rates for $B^+\to\pi^+\pi^0$,
$B^0\to\pi^+\pi^-$ and $B^0\to\pi^0\pi^0$,  as well as their CP-conjugate
counterparts, it is possible to isolate the $I=2$ component and obtain
$\alpha$ with no theoretical uncertainty. Thus, even in the presence of
penguin diagrams, $\sin 2\alpha$ can in principle be extracted from the
decays $B\to\pi\pi$. It must be admitted, however, that this isospin
program is ambitious experimentally. If it cannot be carried out, the error
induced on $\sin 2\alpha$ is of order $|P/T|$, where $P$ ($T$) represents
the penguin (tree) diagram. The ratio $|P/T|$ is difficult to estimate --
it is dominated by hadronic physics. However, one ingredient is the ratio
of the CKM elements of the two contributions: $|V_{tb}^* V_{td} / V_{ub}^*
V_{ud} | \simeq |V_{td}/V_{ub}|$. From our fits, we have determined the
allowed values of $|V_{td}|$ as a function of $|V_{ub}|$. This is shown in
Fig.~\ref{vtdvub} for the ``combined fit". The allowed range for the ratio
of these CKM matrix elements is
\beq
1.2 \leq \left\vert {V_{td}\over V_{ub}} \right\vert \leq 5.8 ~,
\eeq
with the central value close to 3.

It is $\bsbarp\to\rho K_S$ which is most affected by penguins. In fact,
the penguin contribution is probably larger in this process than the tree
contribution. This decay is clearly not dominated by one weak (tree)
amplitude, and thus cannot be used as a clean probe of the angle $\gamma$.
Instead, two other methods have been devised, not involving CP-eigenstate
final states. The CP asymmetry in the decay $\bsbarp\to D_s^\pm K^\mp$ can
be used to extract $\sin^2 \gamma$ \cite{ADK}. Similarly, the CP asymmetry
in  $B^\pm\to\dcp K^\pm$ also measures $\sin^2 \gamma$ \cite{growyler}.
Here, $\dcp$ is a $D^0$ or $\dbar$ which is identified in a CP-eigenstate
mode (e.g.\ $\pi^+\pi^-$, $K^+K^-$, ...).

These CP-violating asymmetries can be expressed straightforwardly in terms
of the CKM parameters $\rho$ and $\eta$. The 95\% C.L.\ constraints on
$\rho$ and $\eta$ found previously can be used to predict the ranges of
$\sin 2\alpha$, $\sin 2\beta$ and $\sin^2 \gamma$ allowed in the standard
model. The allowed ranges which correspond to each of the figures in
Fig.~\ref{rhoeta1}, obtained from Fit 1, are found in Table \ref{cpasym1}.
In this table we have assumed that the angle $\beta$ is measured in
$\bdbarp\to J/\Psi K_S$, and have therefore included the extra minus sign
due to the CP of the final state.

\begin{table}
\hfil
\vbox{\offinterlineskip
\halign{&\vrule#&
 \strut\quad#\hfil\quad\cr
\noalign{\hrule}
height2pt&\omit&&\omit&&\omit&&\omit&\cr
& $\fbd\sqrt{\hat{B}_{B_d}}$ (MeV) && $\sin 2\alpha$ &&
$\sin 2\beta$ && $\sin^2 \gamma$ & \cr
height2pt&\omit&&\omit&&\omit&&\omit&\cr
\noalign{\hrule}
height2pt&\omit&&\omit&&\omit&&\omit&\cr
& $130$ && 0.46 -- 0.88 && 0.21 -- 0.37 && 0.12 -- 0.39  & \cr
& $155$ && 0.75 -- 1.0 && 0.31 -- 0.56 && 0.34 -- 0.92 & \cr
& $180$ && $-$0.59 -- 1.0 && 0.42 -- 0.73 && 0.68 -- 1.0 & \cr
& $205$ && $-$0.96 -- 0.92 && 0.49 -- 0.86 && 0.37 -- 1.0 & \cr
& $230$ && $-$0.98 -- 0.6 && 0.57 -- 0.93 && 0.28 -- 0.97 & \cr
height2pt&\omit&&\omit&&\omit&&\omit&\cr
\noalign{\hrule}}}
\caption{The allowed ranges for the CP asymmetries $\sin 2\alpha$, $\sin
2\beta$ and $\sin^2 \gamma$, corresponding to the constraints on $\rho$ and
$\eta$ shown in Fig.~\protect\ref{rhoeta1}. Values of the coupling constant
$\fbd\protect\sqrt{\hat{B}_{B_d}}$ are stated. We fix $\hat{B}_K=0.8$. The
range for $\sin 2\beta$ includes an additional minus sign due to the CP of
the final state $J/\Psi K_S$.}
\label{cpasym1}
\end{table}

\begin{figure}
\vskip -2.4truein
\centerline{\epsfxsize 7.0 truein \epsfbox {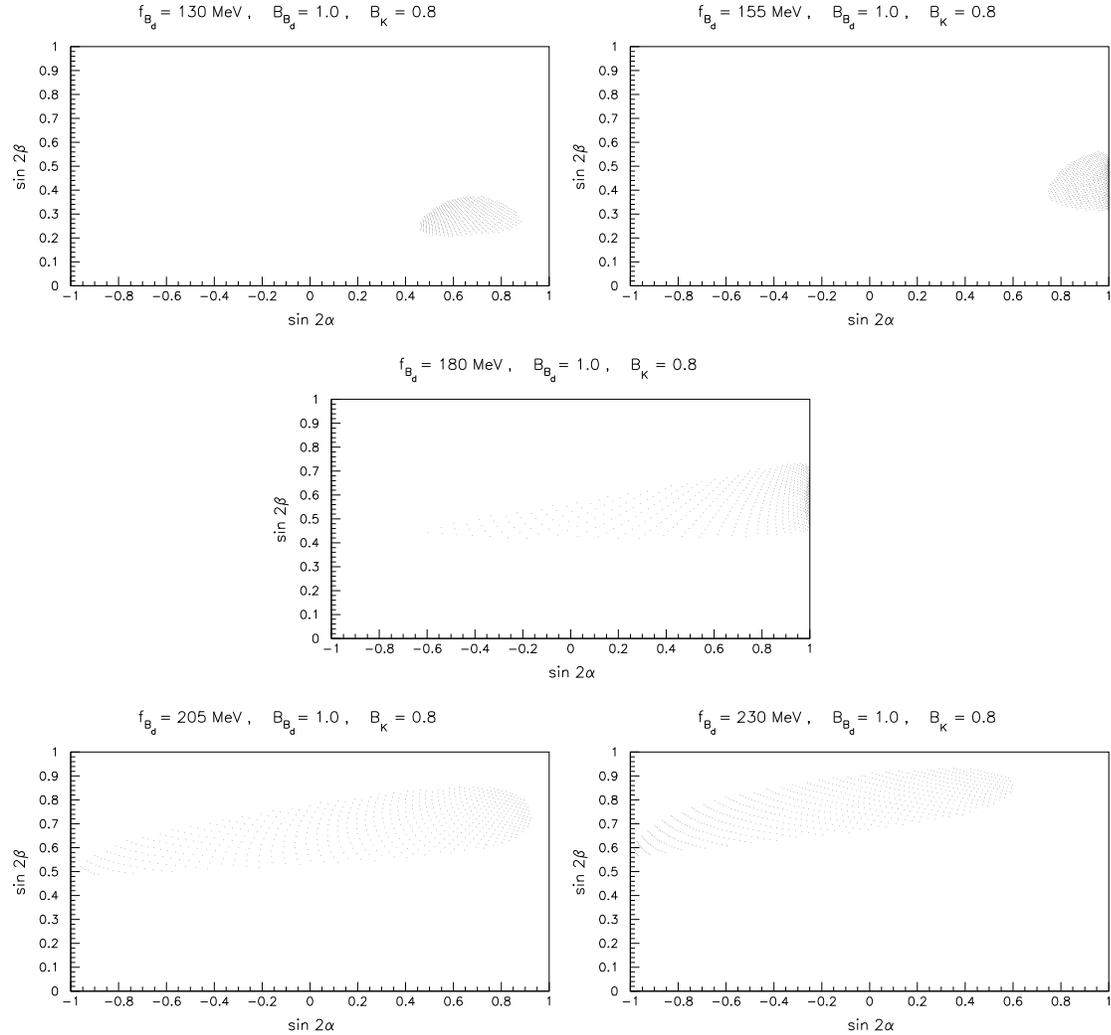}}
\vskip -1.0truein
\caption{Allowed region of the CP asymmetries $\sin 2\alpha$ and $\sin
2\beta$ resulting from the ``experimental fit" of the data for different
values of the coupling constant $\fbd\protect\sqrt{\hat{B}_{B_d}}$
indicated on the figures a) -- e). We fix $\hat{B}_K=0.8$.}
\label{alphabeta1}
\end{figure}

\begin{figure}
\vskip -1.0truein
\centerline{\epsfxsize 3.5 truein \epsfbox {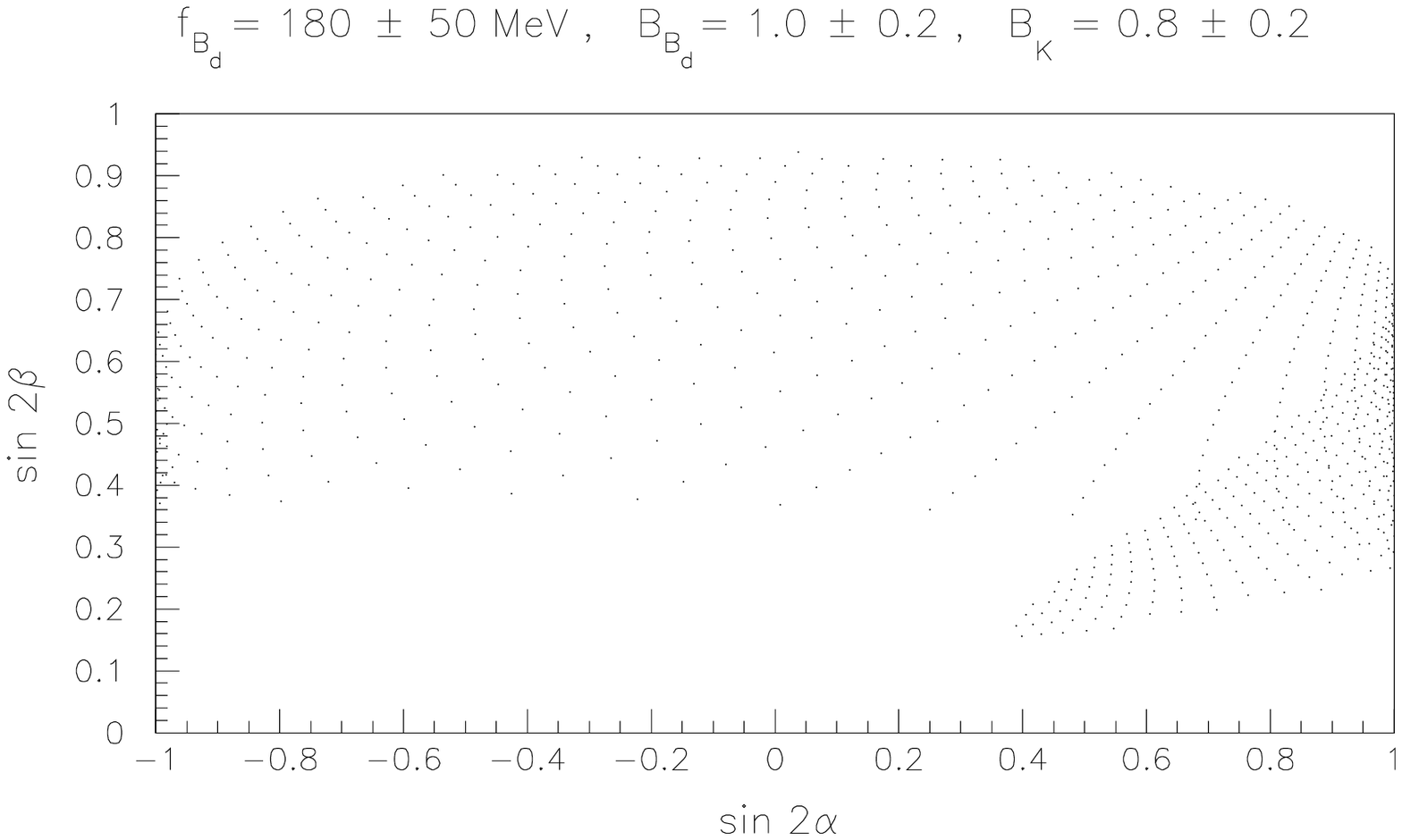}}
\vskip -1.4truein
\caption{Allowed region of the CP asymmetries $\sin 2\alpha$ and $\sin
2\beta$ resulting from the ``combined fit" of the data for the ranges for
$\fbd\protect\sqrt{\hat{B}_{B_d}} $ and $\hat{B}_K$ given in the text.}
\label{alphabeta2}
\end{figure}

Since the CP asymmetries all depend on $\rho$ and $\eta$, the ranges for
$\sin 2\alpha$, $\sin 2\beta$ and $\sin^2 \gamma$ shown in Table
\ref{cpasym1} are correlated. That is, not all values in the ranges are
allowed simultaneously. We illustrate this in Fig.~\ref{alphabeta1},
corresponding to the ``experimental fit" (Fit 1), by showing the region in
$\sin 2\alpha$-$\sin 2\beta$ space allowed by the data, for various values
of $\fbd\sqrt{\hat{B}_{B_d}}$. Given a value for
$\fbd\sqrt{\hat{B}_{B_d}}$, the CP asymmetries are fairly constrained.
However, since there is still considerable uncertainty in the values of the
coupling constants, a more reliable profile of the CP asymmetries at
present is given by our ``combined fit" (Fit 2), where we convolute the
present theoretical and experimental values in their allowed ranges. The
resulting correlation is shown in Fig.~\ref{alphabeta2}. From this figure
one sees that the smallest value of $\sin 2\beta$ occurs in a small region
of parameter space around $\sin 2\alpha\simeq 0.4$-0.6. Excluding this
small tail, one expects the CP-asymmetry in $\bdbarp\to J/\Psi K_S$ to be
at least 30\%.

It may be difficult to extract $\gamma$ using the techniques described
above. First, since $\bsbarp\to D_s^\pm K^\mp$ involves the decay of $B_s$
mesons, such measurements must be done at hadron colliders. At present, it
is still debatable whether this will be possible. Second, the method of
using $B^\pm\to\dcp K^\pm$ to obtain $\gamma$ requires measuring the rate
for $B^+ \to D^0 K^+$. This latter process has an expected branching ratio
of $\lsim O(10^{-6})$, so this too will be hard.

Recently, a new method to measure $\gamma$ was proposed \cite{PRL}. Using a
flavour SU(3) symmetry, along with the neglect of exchange- and
annihilation-type diagrams, it was suggested that $\gamma$ could be found
by measuring rates for the decays $B^+ \to \pi^0 K^+$, $B^+ \to \pi^+ K^0$,
$B^+ \to \pi^+ \pi^0$, and their charge-conjugate processes. The $\pi K$
final states have both $I=1/2$ and $I=3/2$ components. The crucial
ingredient is that the gluon-mediated penguin diagram contributes only to
the $I=1/2$ final state. Thus, a linear combination of the $B^+ \to \pi^0
K^+$ and $B^+ \to \pi^+ K^0$ amplitudes, corresponding to $I = 3/2$ in the
$\pi K$ system, can be related via flavour SU(3) to the purely $I = 2$
amplitude in $B^+ \to \pi^+ \pi^0$, permitting the construction of an
amplitude triangle. The difference in the phase of the $B^+ \to \pi^+
\pi^0$ side and that of the corresponding triangle for $B^-$ decays was
found to be $2 \gamma$. SU(3) breaking can be taken into account by
including a factor $f_K/f_\pi$ in relating $B\to\pi\pi$ decays to the
$B\to\pi K$ decays \cite{GHLR}.

The key assumption is that the penguin is mediated by gluon exchange.
However, there are also electroweak contributions to the penguins
\cite{EWPs}. These electroweak penguins (EWP's) are not constrained to be
isosinglets. Thus, in the presence of EWP's, there is no longer a triangle
relation $B\to \pi K$ and $B\to\pi\pi$ amplitudes \cite{DH}. Indeed,
electroweak penguins can, in principle, even invalidate the isospin
analysis in $B\to \pi\pi$, since the $I=2$ amplitude will include a
contribution from EWP's, and hence will no longer have a well-defined weak
CKM phase. However, theoretical estimates \cite{DH,GHLREWP} show that
electroweak penguins are expected to be relatively unimportant for
$B\to\pi\pi$.

The question of the size of EWP's has therefore become a rather interesting
question, and a number of papers have recently appeared discussing this
issue \cite{EWPsize}. These include both theoretical predictions, as well
as ways of isolating EWP's experimentally. The general consensus is that
EWP's are large enough to invalidate the method of Ref.~\cite{PRL} for
obtaining $\gamma$. However, two new methods making use of the flavour
SU(3) symmetry, and which do not have any problems with electroweak
penguins, have been suggested. Both are rather complicated, making use of
the isospin quadrangle relation among $B\to\pi K$ decays, as well as
$B^+\to\pi^+\pi^0$ plus an additional decay: $B_s \to \eta\pi^0$ in one
case \cite{GHLREWP}, $B^+ \to \eta K^+$ in the other \cite{DH2}. Although
electroweak penguins do not cause problems, SU(3)-breaking effects which
cannot be parametrized simply as a ratio of decay constants are likely to
introduce errors of about 25\% into both methods. It is clear that this is
a subject of great interest at the moment, and work will no doubt continue.


\section{Summary and Outlook}

We summarize our results:

\smallskip

(i) We have presented an update of the CKM unitarity triangle using the
theoretical and experimental improvements in the following quantities:
$\absvcb$, $\vert V_{ub}/V_{cb} \vert$, $\delmd$, $\tau(B_d)$,
${\overline{\mt}}$,
$\hat{\eta}_{cc}$, $\hat{\eta}_{ct}$. The fits can be used to exclude
extreme values of the pseudoscalar coupling constants, with the range
$130~\mbox{MeV} \leq f_{B_d} \sqrt{\hat{B}_{B_d}} \leq 270~\mbox{MeV}$
still allowed for $\hat{B}_K=1$. The lower limit of this range is quite
$\hat{B}_K$-independent, but the upper limit is strongly correlated with
the value chosen for $\hat{B}_K$. For example, for $\hat{B}_K=0.8$ and
$0.6$, $f_{B_d} \sqrt{\hat{B}_{B_d}} \leq 240$ and 210 MeV, respectively,
is required for a good fit. The solutions for $\hat{B}_K = 0.8 \pm 0.2$
are slightly favoured by the data as compared to the lower values. These
numbers are in very comfortable agreement with QCD-based estimates from sum
rules and lattice techniques. The statistical significance of the fit is,
however, not good enough to determine the coupling constant more precisely.
We note that $\hat{B}_K \leq 0.4$ is strongly disfavoured by the data,
since the quality of fit for such values is very poor.

\smallskip

(ii) The newest experimental and theoretical numbers restrict the allowed
CKM unitarity triangle in the $(\rho,\eta)$-space somewhat more than
before. However, the present uncertainties are still enormous -- despite
the new, more accurate experimental data, our knowledge of the unitarity
triangle is still poor. This underscores the importance of measuring
CP-violating rate asymmetries in the $B$ system. Such asymmetries are
largely independent of theoretical hadronic uncertainties, so that their
measurement will allow us to accurately pin down the parameters of the CKM
matrix. Furthermore, unless our knowledge of the pseudoscalar coupling
constants improves considerably, better measurements of such quantities as
$\xd$ will not help much in constraining the unitarity triangle. On this
point, help may come from the experimental front. It may be possible to
measure the parameter $\fbd$, using isospin symmetry, via the
charged-current decay $\bu\to\tau^\pm \nu_\tau$. With $\vert V_{ub}/V_{cb}
\vert =0.08 \pm 0.02$ and $\fbd=180\pm 50~{\rm MeV}$, one gets a branching
ratio $BR(\bu\to\tau^\pm\nu_\tau)=(1.5$--$14.0)\times 10^{-5}$, with
a central value of $5.2\times 10^{-5}$. This
lies in the range of the future LEP and asymmetric $B$-factory experiments,
though at LEP the rate $Z \to B_c X \to \tau^\pm \nu_\tau X$ could be just
as large as $Z \to B^\pm X \to \tau^\pm \nu_\tau X$. Along the same lines,
the prospects for measuring $(\fbd,\fbs)$ in the FCNC leptonic and photonic
decays of $\bd $ and $\bs$ hadrons, $(\bd,\bs)\to\mu^+\mu^-, (\bd,\bs) \to
\gamma\gamma$ in future $B$ physics facilities are not entirely dismal
\cite{ALIINT94}.

\smallskip

(iii) We have determined bounds on the ratio $\vert V_{td}/V_{ts} \vert$
from our fits. For $130~\mbox{MeV} \leq f_{B_d} \sqrt{\hat{B}_{B_d}} \leq
270~\mbox{MeV}$, i.e.\ in the entire allowed domain, at 95 \% C.L. we find
\beq
0.13 \leq \left\vert {V_{td} \over V_{ts}} \right\vert \leq 0.35~.
\eeq
The upper bound from our analysis is more restrictive than the current
experimental upper limit following from the CKM-suppressed radiative
penguin decays $BR(B \to \omega + \gamma )$ and $BR(B \to \rho + \gamma )$,
which at present yield at 90\% C.L. \cite{cleotdul}
\beq
\left\vert {V_{td} \over V_{ts}} \right\vert \leq 0.64 - 0.75~,
\eeq
depending on the model used for the SU(3)-breaking in the relevant form
factors \cite{SU3ff}. Long-distance effects in the decay $B^\pm \to
\rho^\pm + \gamma$ may introduce theoretical uncertainties comparable to
those in the SU(3)-breaking part but the corresponding effects in the
decays $B^0 \to (\rho^0,\omega) +\gamma$ are expected to be very small
\cite{AB95}. Furthermore, the upper bound is now as good as that obtained
from unitarity, which gives $0.08 \leq \vert V_{td}/V_{ts} \vert \leq
0.36$, but the lower bound from our fit is more restrictive.

\smallskip

(iv) Using the measured value of $\mt$, we find
\begin{equation}
\xs = \left(20.7 \pm 4.5\right)\frac{\fbbs}{(230 ~\mbox{MeV})^2}~.
\end{equation}
Taking $f_{B_s}\sqrt{\hat{B}_{B_s}}= 230$ (the central value of lattice-QCD
estimates), and allowing the coefficient to vary by $\pm 2\sigma$, this
gives
\begin{equation}
11.7 \leq \xs \leq 29.7~.
\end{equation}
No reliable confidence level can be assigned to this range -- all that one
can conclude is that the SM predicts large values for $\xs$, most of which
lie above the ALEPH 95\% C.L. lower limit of $\xs > 8.8$.

\smallskip

(v) The ranges for the CP-violating rate asymmetries parametrized by $\sin
2\alpha$, $\sin 2\beta$ and and $\sin^2 \gamma$ are determined at 95\% C.L.
to be
\begin{eqnarray}
&~& -1.0 \leq \sin 2\alpha \le 1.0~, \nonumber \\
&~& 0.21 \leq \sin 2\beta \le 0.93~, \\
&~& 0.12 \leq \sin^2 \gamma \le 1.0~. \nonumber
\end{eqnarray}
(For $\sin 2\alpha < 0.4$, we find $\sin 2\beta \ge 0.3$.) Electroweak
penguins may play a significant role in some methods of extracting
$\gamma$. Their magnitude, relative to the tree contribution, is therefore
of some importance. One factor in determining this relative size is the
ratio of CKM matrix elements $\vert V_{td}/V_{ub} \vert$. We find
\beq
1.2 \leq \left\vert {V_{td}\over V_{ub}} \right\vert \leq 5.8 ~.
\eeq

\bigskip
\noindent
{\bf Acknowledgements}:
\bigskip

We thank Viladimir Braun, Roger Forty, Christoph Greub, Matthias Neubert,
Olivier Schneider, Vivek Sharma, Tomasz Skwarnicki, Ed Thorndike, Pippa
Wells and Sau Lan Wu for very helpful discussions. A.A. thanks the
hospitality of the organizing committee for the 6$^{th}$ symposium on heavy
flavour physics in Pisa.



\end{document}